# Cavity electromagnetically induced transparency and all-optical switching using ion Coulomb crystals

Magnus Albert[1], Aurélien Dantan[1] & Michael Drewsen[1]

[1]*QUANTOP – Danish National Research Foundation Center for Quantum Optics, Department of Physics and Astronomy, Aarhus University, DK-8000 Aarhus, Denmark*

**The control of one light field by another, ultimately at the single photon level[1-6], is a challenging task which has numerous interesting applications within nonlinear optics[4,5] and quantum information science[6,7]. Due to the extremely weak direct interactions between optical photons in vacuum, this type of control can in practice only be achieved through highly nonlinear interactions within a medium[1-9].**

**Electromagnetic induced transparency (EIT)[1,5] constitutes one such means to obtain the extremely strong nonlinear coupling needed to facilitate interactions between two faint light fields[2-6,8-11].**

**Here, we demonstrate for the first time EIT as well as all-optical EIT-based light switching using ion Coulomb crystals situated in an optical cavity. Unprecedented narrow cavity EIT feature widths down to a few kHz and a change from essentially full transmission to full absorption of the probe field within a window of only ~100 kHz are achieved. By applying a weak switching field, we furthermore demonstrate nearly perfect switching of the transmission of the probe field. These results represent important milestones for future realizations of quantum information processing devices, such as high-efficiency quantum memories[12,13], single-photon transistors[14,15] and single-photon gates[4,6,8].**

Electromagnetically induced transparency is a quantum interference phenomenon appearing when two electromagnetic fields excite resonantly two different transitions sharing a common state[1,5]. An intense control field addressing one of the transitions can lead to a dramatic change in the linear absorption and dispersion properties of a weak probe field being resonant with the other. The application of a control field can, e.g., lead to full transmission of a probe field propagating in an otherwise optically dense medium[1,5]. In the past, EIT has successfully been applied for slowing and stopping pulses of light[16-18], as well as for storing and retrieving single-photons in hot and cold atomic gasses[7,19,20].

The refractive index around an EIT resonance can furthermore lead to extremely large nonlinear couplings between the probe field and an additional third field coupling the non-common state of the control field transition to a fourth state[9-11]. Ultimately, in a cavity setup, where the light-matter coupling is greatly enhanced, a "photon blockade" scenario in which the transmission of a single photon is coherently controlled by another should be achievable[2,3]. As an alternative to the application of the anharmonicity of the Jaynes-Cummings eigenenergies for a single atom in a cavity[14], such strong effective photon-photon interactions could serve as the basis for realizing fundamental quantum devices, such as single-photon transistors[14,15] and single-photon quantum gates[4,6,8], and could allow for the observation of novel quantum phase transitions for light[21].

Recently, cavity EIT features have been observed with a medium consisting either of a few [22,23] or larger ensembles of neutral atoms[24,25]. While the signature of EIT in free space has as well been observed with a single ion[26], ensembles of cold ions, placed in an optical cavity and in the so-called collective strong coupling regime[27], are promising for investigating EIT-related phenomena in optically dense media.

Here, we report on the first observations of cavity EIT using an ensemble of cold ions in the form of a Coulomb crystal of $^{40}Ca^+$ ions enclosed in a moderately high-finesse (~3000) linear cavity. The cavity EIT signals are obtained in a novel configuration where both the probe and the control fields are injected into the same spatial cavity mode (Fig. 1a). The weak probe field (single photon level) and the more intense control field, which originate from the same laser, are both tuned to the 3d $^2D_{3/2}$-4p $^2P_{1/2}$ transition (866 nm), but with opposite circular polarizations to address the $m_J$=+3/2 and $m_J$=-1/2 Zeeman sub-states of the $3d^2D_{3/2}$ level, respectively (Fig. 1b). Due to the long coherence times of the $m_J$=+3/2 and $m_J$=-1/2 Zeeman sub-states, complete transparency in the otherwise fully absorbing medium can be achieved with a control field corresponding to ~500 intracavity photons. The typical transparency window is only a few tens of kHz, i.e. one to two orders of magnitude narrower than observed in cavity EIT experiments with neutral atoms[22-25]. In addition, we make use of these narrow transparency windows to implement a variant of the four-level cavity EIT-blockade scheme of Ref. 2, in which the transmission of the probe photons (866nm) is coherently controlled by another weak field (850 nm) tuned to the $^2D_{3/2}$-4p $^2P_{3/2}$ transition, and injected into the same spatial cavity mode (Figs. 1a,c). The observations are found to be in excellent agreement with a simple theoretical model.

A sketch of the essential optical components of the experimental setup is presented in Fig. 1a. The ion Coulomb crystals are produced in a linear radiofrequency trap incorporating an 11.8 mm-long optical cavity[27]. The ions are Doppler laser-cooled to temperatures in the 10 mK range and prepared by optical pumping in the $m_J$=+3/2 substate of the $3d^2D_{3/2}$ level with ~97% efficiency[27]. The trapping potentials are chosen such that the ion Coulomb crystals, spheroidal with respect to the cavity axis, have a typical length of 1-2 mm, radii of 200-300 μm and densities of $5.6 \times 10^8$ cm$^{-3}$. To probe the spectrum of the coupled system consisting of the crystal and the cavity, a $\sigma^-$-circularly polarized probe pulse tuned around the resonance of the $3d^2D_{3/2}$, $m_J$=+3/2 - $4p^2P_{1/2}$, $m_J$=+1/2 transition is injected into the cavity through a relative high-transmitting mirror ($T_H$=1500 ppm) with an intensity such that the average intracavity photon number is less than one at any time. The spectral response is determined through detection of the reflected photons by an avalanche photodiode. To observe an EIT signal, an additional $\sigma^+$-circularly polarized control pulse tuned to resonance with the $3d^2D_{3/2}$, $m_J=-1/2$ - $4p^2P_{1/2}$, $m_J=+1/2$ transition is injected into the cavity through the low-transmitting mirror ($T_L$=4 ppm). In all the experiments, the cavity resonance frequency is kept fixed at the $3d^2D_{3/2}$, $m_J=+3/2$ - $4p^2P_{1/2}$, $m_J=+1/2$ resonance.

Typical measured reflection spectra for the probe, as a function of the probe detuning from atomic resonance $\Delta$, are shown in Fig. 2**a** for three cases. For a bare cavity, i.e. no crystal present, the spectrum is Lorentzian with a half-width corresponding to a cavity field decay rate $\kappa= 2\pi\times(2.2\pm0.1)$ MHz, as expected from the mirror transmissions and an additional absorption loss of ~650 ppm. In a situation where the cavity contains a crystal with ~680 ions effectively interacting with the probe field, the collective strong coupling regime, where the collective coupling rate[27] $g_N = 2\pi\times(13.8\pm0.1)$ MHz supersedes both the spontaneous emission rate $\gamma=2\pi\times(12.6\pm0.5)$ MHz and the cavity field decay rate $\kappa$, has been reached. Here, one observes the characteristic single-photon Rabi splitting. Finally, when both the control and the probe fields are injected into the cavity, a narrow EIT feature is observed around the two-photon resonance condition. The half-width-half-maximum (HWHM) of the central EIT feature is only 47.5$\pm$2.4 kHz, i.e. about 50 times narrower than the bare cavity half-width, and at two-photon resonance, the atomic transparency is substantially increased from only 2% without the coupling field to 84%. A fit to this spectrum according to a theoretical model (see Methods) yields a collective coupling rate $g_N=2\pi\times(13.5\pm0.2)$ MHz and a control field intracavity Rabi frequency $\Omega_c=2\pi\times(4.1\pm0.1)$ MHz. Figure 2**b** presents a zoom of the spectra around the two-photon resonance. The non-Lorentzian shape of the EIT resonance can be well-explained by taking into account the fact that both control and probe fields are coupled into the same spatial ($TEM_{00}$) cavity mode, which have a radially varying intensity profile (see Methods).

To investigate the cavity EIT feature in more detail, EIT spectra similar to those presented in Fig. 1 have been recorded with different intensities of the control pulse. In Fig. 2**c**, a selection of such spectra is presented in the case of a crystal with ~980 effectively interacting ions (corresponding to a cooperativity $C = g_N^2 / 2\kappa\gamma \simeq 5.4$) and for control field Rabi frequencies $\Omega_c = 2\pi \times (1.3, 3.0, 6.6, 8.6)$ MHz. The solid lines in this figure originate from a global best fit of the model presented in the Method section to all 11 obtained spectra with fixed values of the control field Rabi frequencies and with the phenomenological decoherence rate $\gamma_0$ of the $3d^2D_{3/2}, m_J=-1/2 - 3d^2D_{3/2}, m_J=+3/2$ transition as the only free parameter. From these fit a decoherence rate $\gamma_0 = 2\pi \times (0.6 \pm 0.1)$ kHz was obtained. Taking into account the uncertainties in the control field Rabi frequencies gives a value of $\gamma_0$ in the range $2\pi \times (0.3 - 1.1)$ kHz, in good agreement with the Zeeman decoherence rate found in our previous studies of collective strong coupling[27].

Figure 2**d** shows the maximal atomic transparency (definition: see Methods) as well as the width of the EIT feature obtained from fits to the experimental recorded spectra using the previously determined value range of $\gamma_0$. While the width of the EIT feature is increasing linearly with $\Omega_c^2$, or equivalently with the control pulse intensity, the maximal transparency has a saturating-like behaviour. As a consequence, it is possible to change the atomic transparency of a probe pulse from ~1% to more than 90% by changing its detuning by as little as ~100 kHz.

This situation is an excellent starting point to further study the prospect of an all-optical switching of the cavity probe field transmission by injecting yet another weak *switching* field into the cavity, as sketched in Figs. 1**a,c**. In the experiments, the switching field is tuned as close as possible to the $3d^2D_{3/2}$- $4p^2P_{3/2}$ transition (850 nm), while at the same time being resonant with one of the TEM$_{00}$ cavity modes (i.e. a detuning of -4.3 GHz). The field is injected through the low-transmitting mirror ($T_L$), and $\sigma^+$-circularly polarized in order to only couple to the initially empty $3d^2D_{3/2}$, $m_J=-1/2$ sub-state.

The main effect of the far-off resonant switching field is to induce a light-shift[9] of the $3d^2D_{3/2}$, $m_J=-1/2$ sub-state coupled by the control field, and hence to change the condition for the two-photon EIT resonance. By turning on and off the switching field the probe field will "see" the two-photon EIT resonance shifted by exactly this light-shift. Due to the very narrow EIT resonances obtained, switching of the transmission of the probe field can be controlled by a rather weakly coupled switching field.

This fact is illustrated in Fig. 3**a**, where the probe reflectivity spectrum around resonance is shown for different switching field input powers, and for a crystal with effectively ~920 ions interacting with all three cavity fields. As the switching field intensity (or equivalently $\Omega_c^2$) is increased, the EIT resonance is red-shifted and the probe absorption level at the unperturbed EIT resonance ($\Delta=0$) gradually increases to the fully absorbing situation (no EIT). The asymmetric shapes of the shifted spectra arise from the transverse Gaussian intensity distribution of the fields in the cavity (see Methods). In Fig. 3**b**, the atomic transparency of the crystal for the probe tuned to $\Delta=0$ and the shift of the EIT resonance are shown as a function of the switching field intensity ($\Omega_s^2$). From this figure one can deduce a nearly linear shift of the EIT resonance of ~0.7 Hz per switching field photon, and that typically ~80,000 inter-cavity photons are needed for a nearly perfect switching (better than 90%) in the current setup. A simple extrapolation of the measured shifts to a situation where the probe (866 nm) and switching (850 nm) fields are both resonant with a cavity mode and with the $3d^2D_{3/2}$- $4p^2P_{1/2}$ and $3d^2D_{3/2}$- $4p^2P_{3/2}$ transitions, respectively, suggests that switching by a few intracavity photons should be feasible with our present experimental setup after an adjustment of the cavity length.

The observed cavity EIT and photon switching with ion Coulomb crystals are extremely promising for the realization of both high-efficiency *and* long-lived quantum memories[12,13] as well as for further studies of single-photon nonlinear effects[2,3]. In addition, owing to their dilute solid-like nature, ion Coulomb crystals possess unique properties (e.g. uniform density and tunable vibrational mode spectrum) for realizing multimode quantum interfaces by exploiting their spatial[28] or motional degrees of freedom[29]. Finally, the results are highly interesting for investigations of EIT-based optomechanical phenomena in analogy with recent studies based on usual solids[30].

**Methods. Theoretical model: EIT and all-optical switching susceptibilities**

Assuming that the cavity resonance frequency is equal with the frequency $\omega_{at}$ of the $3d^2D_{3/2}$, $m_J=+3/2$ - $4p^2P_{1/2}$, $m_J=+1/2$ transition and using the standard linear approximations for the field in a high-finesse cavity, the steady state mean value of the intracavity probe field amplitude $\langle a \rangle$ is given by $\langle a \rangle = \frac{\sqrt{2\kappa_{T_H}/\tau}\langle a^{in}\rangle}{\kappa - i\Delta - i\chi}$, where $\kappa$ is the total cavity field decay rate, $\kappa_{T_H}$ is the cavity field decay rate specifically due to transmission losses through mirror $T_H$, $\Delta=\omega_p-\omega_{at}$ the probe field detuning, $\omega_p$ the probe field frequency, $\tau$ the cavity round-trip time, $\langle a^{in} \rangle$ is the injected probe field mean value and $\chi$ is the optical susceptibility of the ion crystal. Using this formula the probe reflectivity is then generally given by $R = \left|\frac{2\kappa_{T_H}}{\kappa - i\Delta - i\chi} - 1\right|^2$. The atomic transparency for a resonant probe and cavity ($\Delta=0$) is defined by the ratio of the transmission of the cavity containing the medium T to that of the empty cavity $T_0$: $\frac{T}{T_0} = \frac{\kappa^2}{(\kappa + \text{Im}[\chi])^2 + \text{Re}[\chi]^2}$. For an interaction with a resonant probe field only ($\text{Re}[\chi]=0$), $\text{Im}[\chi]/\kappa = g_N^2/(\kappa\gamma) = 2C$ is given by (twice) the cooperativity parameter.

The susceptibility in the EIT configuration can be calculated by solving the standard optical Bloch equations for an ensemble of three-level $\Lambda$ atoms interacting with two cavity fields. Since both probe and control fields are in the TEM$_{00}$ mode of the cavity, their spatial profile over the crystal volume should be taken into account. Since the temperature of the ions is in the ~10 mK range, during the time for the EIT condition to reach steady state, the ions probe any field variations along standing-wave structure of the field[31-33], and hence an average *longitudinal* field amplitude can be assumed. Furthermore, since the lengths of the crystals are much shorter than the Rayleigh length of the cavity mode, the susceptibility can be obtained by integrating the standard linear susceptibility across the *transverse* Gaussian profile of the modes. To first order in the probe field and for a control field tuned to atomic resonance and with Rabi frequency $\Omega_c$ (maximum amplitude at the centre of the cavity mode), the EIT susceptibility is given by $\chi_{EIT} = \frac{ig_N^2}{\gamma - i\Delta} \frac{\ln(1+\Theta)}{\Theta}$, where $\Theta = \frac{\Omega_c^2/2}{(\gamma - i\Delta)(\gamma_0 - i\Delta)}$, $g_N$ is the collective coupling rate[27], and $\gamma_0$ is a phenomenological decay rate for the coherence between the 3d$^2$D$_{3/2}$, m$_J$=-1/2 and m$_J$=+3/2 sub-states.

The susceptibility in the all-optical switching situation can be calculated in a similar fashion, but using the 4-level schemes shown in Fig. 1**c.** In a perturbative treatment of the probe field, this yields a nonlinear susceptibility $\chi_{SW} = \frac{ig_N^2}{\gamma - i\Delta}\left[\frac{\Theta \ln(1+\Theta+\Theta_s)}{(\Theta+\Theta_s)^2} + \frac{\Theta_s}{\Theta+\Theta_s}\right]$, with $\Theta_s = \frac{\Omega_s^2}{(\gamma_s - i\Delta_s)(\gamma_0 - i\Delta)}$ where $\Omega_s$ and $\Delta_s$ are the switching field Rabi frequency and detuning, respectively, and $\gamma_s$ is the decay of the 3d$^2$D$_{3/2}$, m$_J$=+3/2 - 4p$^2$P$_{3/2}$, m$_J$=+3/2 coherence.

1. Harris, S. E. Electromagnetically Induced Transparency. Phys. Today **50**, 36-42 (1997).

2. Imamoglu, A., Schmidt, H., Woods, G. & Deutsch, M. Strongly interacting photons in a nonlinear cavity. Phys. Rev. Lett. **79**, 1467-1470 (1997).

3. Grangier, P., Walls, D. F. & Gheri, K. M. Comment on "Strongly interacting photons in a nonlinear cavity", Phys. Rev. Lett. **81**, 2833-2833 (1998).

4. Harris, S. E. & Hau, L. V. Nonlinear optics at low light levels. Phys. Rev. Lett. **82**, 4611-4614 (1999).

5. Fleischhauer, M., Imamoglu, A. & Marangos, J. P. Electromagnetically induced transparency: Optics in coherent media. Rev. Mod. Phys. **77**, 633-673 (2005).

6. Lukin M. D. & Imamoglu, A. Controlling photons using electromagnetically induced transparency. Nature **413**, 273-276 (2001).

7. Kimble, H. J. The quantum internet. Nature **453**, 1023-1030 (2008).

8. Ottaviani, C., Vitali, D., Artoni, M., Cataliotti, F. & Tombesi, P. Polarization qubit phase gate in driven atomic media. Phys. Rev. Lett. **90**, 197902 (2003).

9. Schmidt, H. & Imamoglu, A. Giant Kerr nonlinearities obtained by electromagnetically induced transparency. Opt. Lett. **21**, 1936-1938 (1996).

10. Kang, H. & Zhu, Y. Observation of large Kerr nonlinearity at low light intensities. Phys. Rev. Lett. **91**, 093601 (2003).

11. Braje, D. A., Balic, V., Yin, G. Y. & Harris, S. E. Low-light-level nonlinear optics with slow light. Phys. Rev. A **68**, 041801(R) (2003).

12. Lukin, M. D., Yelin, S. F. & Fleischhauer, M. Entanglement of atomic ensembles by trapping correlated photon states. Phys. Rev. Lett. **84**, 4232-4235 (2000).

13. Dantan, A. & Pinard, M. Quantum-state transfer between fields and atoms in electromagnetically induced transparency. Phys. Rev. A **69**, 043810 (2004).

14. Birnbaum, K. M., Boca, A., Miller, R., Boozer, A. D., Northup, T. E. & Kimble, H. J. Photon blockade in an optical cavity with one trapped atom. Nature **436**, 87-90 (2005).

15. Hwang, J., Pototschnig, M., Lettow, R., Zumofen, G., Renn, A., Götzinger, S. & Sandoghdar, V. A single molecule optical transistor. Nature **460**, 76-80 (2009).

16. Hau, L. V., Harris, S. E., Dutton, Z. & Behroozi, C. H. Light speed reduction to 17 meters per second in an ultracold atomic gas. Nature **397**, 594-598 (1999).

17. Kash, M. M. *et al.* Ultraslow group velocity and enhanced nonlinear effects in a coherently driven hot atomic gas. Phys. Rev. Lett. **82**, 05229-5232 (1999).

18. Liu, C., Dutton, Z., Behroozi, C. H. & Hau, L. V. Observation of coherent optical information storage in an atomic medium using halted light pulses. Nature **409**, 490-493 (2001).

19. Chanelière, T., Matsukevich, D. N., Jenkins, S. D., Lan, S.-Y., Kennedy, T. A. B. & Kuzmich, A. Storage and retrieval of single photons transmitted between remote quantum memories. Nature **438**, 833-836 (2005).

20. Eisaman, M. D., André, A., Massou, F., Fleischhauer, M., Zibrov, A. S. & Lukin, M. D. Electromagnetically induced transparency with tunable single-photon pulses. Nature **438**, 837-841 (2005).

21. Hartmann, M. J., Brandão, F. G. S. L. & Plenio, M. B. Strongly interacting polaritons in coupled arrays of cavities. Nature Phys. **2**, 849-855 (2006).

22. Mücke, M., Figueroa, E., Bochmann, J., Hahn, C., Murr, K., Ritter, S., Villas-Boas, C. J. & Rempe, G. Electromagnetically induced transparency with single atoms in cavity. Nature **465**, 755-758 (2010).

23. Kampschulte, T., Alt, W., Brakhane, S., Eckstein, M., Reimann, R., Widera, A. & Meschede, D. Optical control of the refractive index of a single atom. Phys. Rev. Lett. **105**, 153603 (2010).

24. Hernandez, G. Zhang, J. & Zhu, Y. Vacuum Rabi splitting and intracavity dark state in a cavity-atom system. Phys. Rev. A **76**, 053814 (2007).

25. Wu, H., Gea-Banacloche, J. & Xiao, M. Observation of intracavity electromagnetically induced transparency and polariton resonances in a Doppler-broadened medium. Phys. Rev. Lett. **100**, 173602 (2008).

26. Slodicka, L., Hétet, G., Gerber, S., Hennrich, M. & Blatt, R. Electromagnetically induced transparency from a single atom in free space. Phys. Rev. Lett. **105**, 153604 (2010).

27. Herskind, P. F., Dantan, A., Marler, J. P., Albert, M. & Drewsen, M. Realization of collective strong coupling with ion Coulomb crystals in an optical cavity. Nature Phys. **5**, 494-498 (2009).

28. Dantan, A., Albert, M., Marler, J. P., Herskind, P. F. & Drewsen, M. Large ion Coulomb crystals: A near-ideal medium for coupling optical cavity modes to matter. Phys. Rev. A **80**, 041802(R) (2009).

29. Dantan, A., Marler, J. P., Albert, M., Guénot, D. & Drewsen, M. Noninvasive vibrational mode spectroscopy of ion Coulomb crystals through resonant collective coupling to an optical cavity field. Phys. Rev. Lett. **105**, 103001 (2010).

30. Weis, S., Rivière, R., Deléglise, S., Gavartin, E., Arcizet, O., Schliesser, A. & Kippenberg, T. J. Optomechanically induced transparency. Science **330**, 1520-1523 (2010).

31. Zimmer, F. E., André, A., Lukin, M. D. & Fleischhauer, M. Coherent control of stationary light pulses. Opt. Comm. **264**, 441-453 (2006).

32. Lin, Y.-W. *et al.* Stationary light pulses in cold atomic media and without Bragg gratings. Phys. Rev. Lett. **102**, 213601 (2009).

33. Wu, J.-H., Artoni, M. & La Rocca, G. C. Stationary light pulses in cold thermal atomic clouds. Phys. Rev. A **82**, 013807 (2010).

We are grateful to Joan Marler for her help at an early stage of these experiments and acknowledge financial support from the Calsberg foundation, the Danish Natural Science Research Council through the ESF EuroQUAM project CMMC and the EU FP7 PICC project. Correspondence and requests for materials should be addressed to M.D. ([drewsen@phys.au.dk](mailto:drewsen@phys.au.dk)). All authors have substantially contributed to this work.

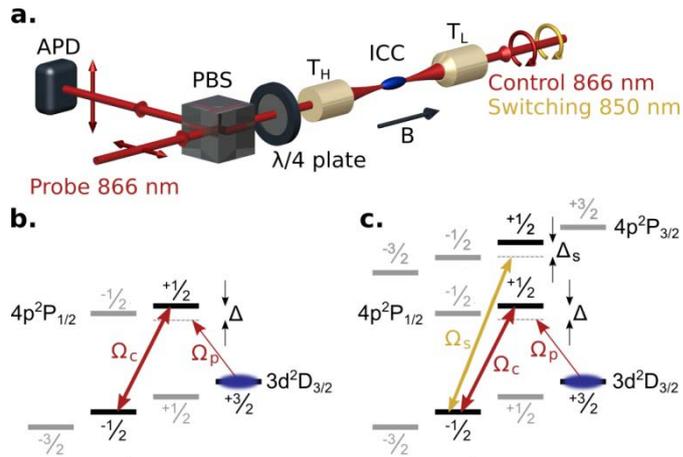

**Figure 1 | Optical setup and level schemes. a,** Schematic of the essential optical parts of the experiment. Using a polarizing beamsplitter (PBS) and a quarter-wave plate, a $\sigma^-$-circularly polarized 866 nm probe field is injected into a linear cavity through mirror $T_H$, where it interacts with an Ion Coulomb Crystal (ICC). The probe light reflected by the cavity is detected by an Avalanche PhotoDiode (APD). In the EIT experiments, a $\sigma^+$-circularly polarized 866 nm control field is injected into the cavity through mirror $T_L$. In the all-optical switching experiments, an additional weak $\sigma^+$-circularly polarized 850 nm field injected through mirror $T_L$. **b,c,** Level schemes applied for (**b**) cavity EIT and (**c**) all-optical switching.

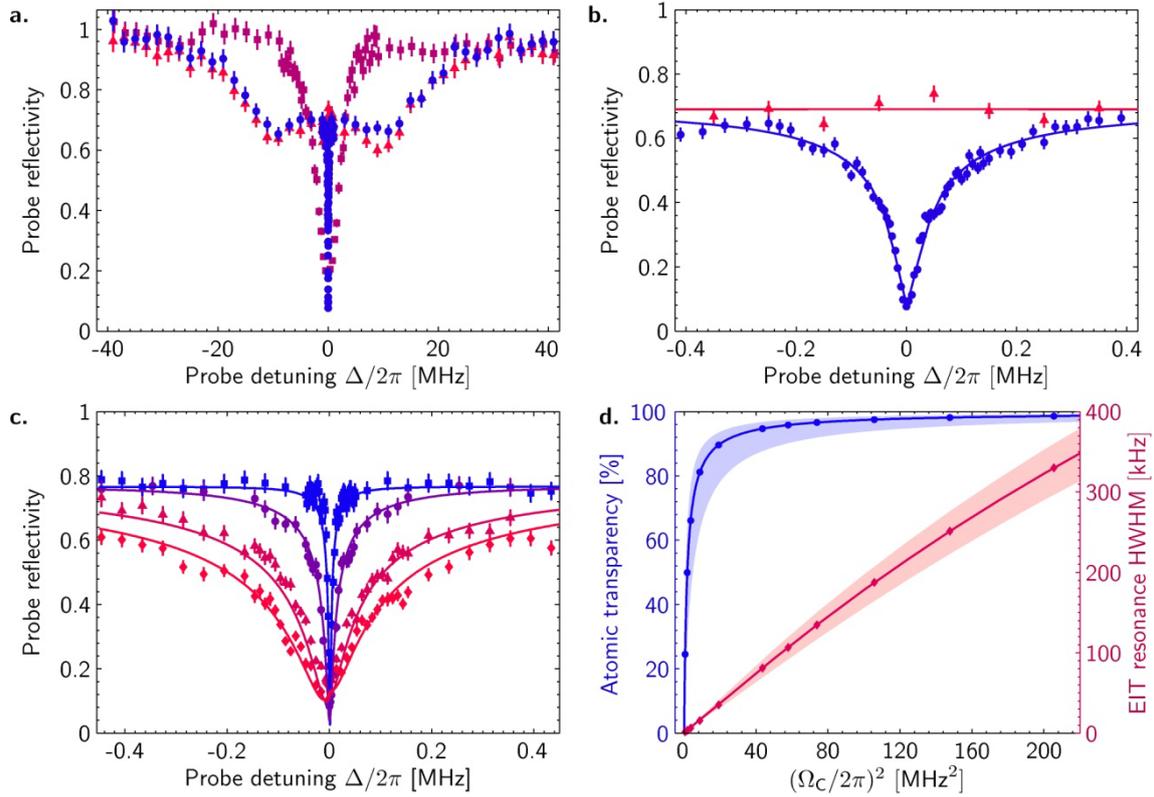

**Figure 2 | Cavity EIT. a,** Probe reflectivity spectrum for an empty cavity (■), as well as for a cavity containing an ion Coulomb crystal (cooperativity C~3.4) while a control field [$\Omega_c=2\pi\times4.1$MHz] is present (●) or not (▲). **b,** Zoom of the spectrum around the EIT resonance. The solid lines are fits based on the theoretical model (see Methods). **c,** EIT spectra obtained for various control field Rabi frequencies $\Omega_c=2\pi$ [1.3 (■), 3.0 (●), 6.6 (▲), 8.6 (♦)] MHz and a cooperativity C~5.4. The solid lines are theoretical fits (See main text). **d,** Calculated atomic transparency (blue) on two-photon EIT resonance and half-widths of the EIT resonance (red) as a function of $\Omega_c^2$. The atomic transparency and EIT widths are obtained from a global fit to 11 spectra similar to the ones shown in **c**. The shaded areas show the estimated error due to the uncertainty in the calibration of $\Omega_c$. The points (●,♦) indicate the value of $\Omega_c$ used in the experiments.

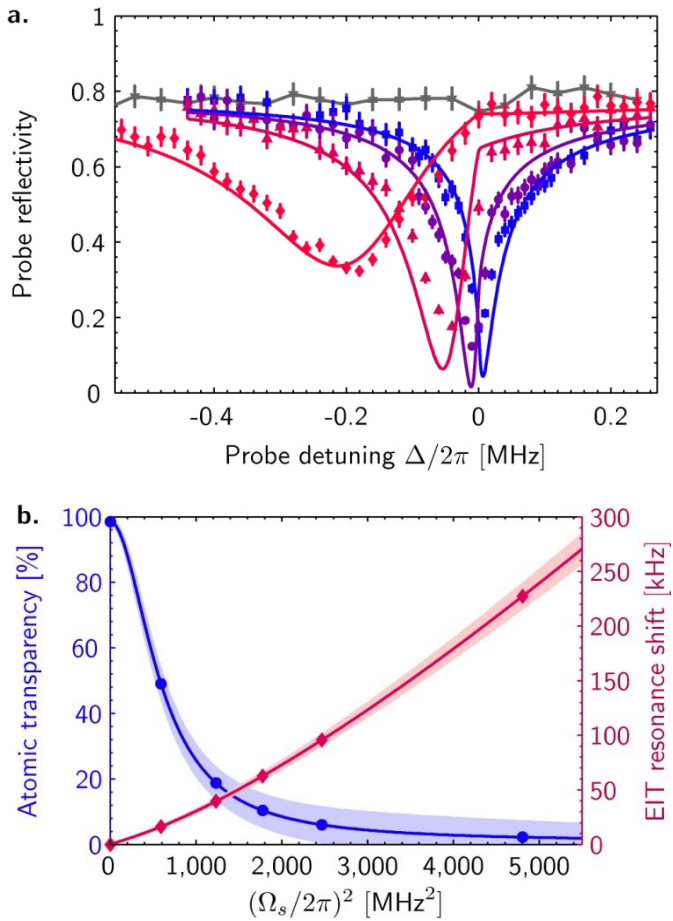

**Figure 3 | All-optical switching. a,** Probe reflectivity spectra for different switching field Rabi frequencies $\Omega_s=2\pi$ [ 0 (■), 24 (●), 42 (▲), 70 (♦)] MHz, and a constant control field Rabi frequency $\Omega_c=2\pi\times 4.2$ MHz and cooperativity C~5.1. For reference, the reflectivity of the probe in presence of the switching field of $\Omega_s=2\pi\times 70$ MHz, but without the control field ($\Omega_c=0$) is also shown (▼). **b,** Atomic transparency of the crystal for the probe tuned to $\Delta=0$ (blue) and the shift of the EIT resonance (red) are shown as a function of the switching field intensity ($\Omega_s^2$). The atomic transparency and shift are calculated from fits based on the model (Methods). The shaded areas show the error due to the uncertainty in the calibration of $\Omega_s$. The points (●,♦) indicate the value of $\Omega_s$ used in the experiments.